\documentclass[global,twocolumn]{svjour}

\usepackage{latexsym}
\usepackage{graphics}
\usepackage{graphicx}
\usepackage{amsmath}
\usepackage{amssymb}

\newcommand{\sgn}{\mathop{\mathrm{sgn}}\nolimits}

\journalname{ApplPhysB}

\begin{document}
\sloppy

\title{Optical polarizabilities of large molecules measured in near-field interferometry}

\author{Lucia Hackerm\"uller\inst{1} \thanks{present address:
Johannes-Gutenberg-Universit\"at Mainz, Institut f\"ur Physik,
Staudingerweg 7, D-55120 Mainz, Germany} \and Klaus
Hornberger\inst{2} \and Stefan Gerlich\inst{1} \and Michael
Gring\inst{1} \and Hendrik Ulbricht\inst{1} \and and Markus
Arndt\inst{1}}

\institute{Faculty of Physics, University of Vienna,
Boltzmanngasse 5, A-1090 Vienna, Austria \\
\and Arnold Sommerfeld Center for Theoretical Physics,
Ludwig-Maximilians-Universit\"{a}t, Theresienstr. 37, D-80333 Munich,
Germany} 

\date{Sent: \today , Received: date / Revised version: date}
\offprints{markus.arndt@univie.ac.at}

\maketitle

\begin{abstract}
We discuss a novel application of matter wave interferometry to
characterize the scalar optical polarizability of molecules at
532\,nm. The interferometer presented here consists of two
material absorptive gratings and one central optical phase
grating. The interaction of the molecules with the standing light
wave is determined by the optical dipole force and is therefore
directly dependent on the molecular polarizability at the
wavelength of the diffracting laser light. By comparing the
observed matter-wave interference contrast with a theoretical
model for several intensities of the standing light wave and
molecular velocities we can infer the polarizability in this
first proof-of-principle experiment for the fullerenes C$_{60}$
and C$_{70}$ and we find a good agreement with literature values.
\end{abstract}


\maketitle

\section{Introduction}\label{sec:intro}
Matter-wave experiments with neutrons, electrons and atoms are
well-established tools for the investigation of fundamental
physical
concepts~\cite{Shull1969a,Mollenstedt1959a,Carnal1991a,Keith1991a},
as well as for innovative and practical measurement applications
~\cite{Rauch2000a,Price1996a,Tonomura1987a,Cronin2007a}.
Coherence experiments with complex molecules are still relatively
young~\cite{Arndt1999a}. But also here interesting applications
have been identified, such as for instance in measurements of
molecular {\em static} polarizabilities
\cite{Berninger2007a,Deachapunya2007a}.

{\em Static} polarizabilities and permanent dipole moments can be
predicted by ab-initio and semi-classical methods
\cite{Deachapunya2007a,Gaussian2003a} and they can be determined
in molecular beam deflection
experiments~\cite{Bonin1997a,Antoine1999a}. All these methods may reach
an accuracy of better than ten percent but they are also
increasingly demanding with increasing particle mass and
complexity.

Atomic Mach-Zehnder interferometry has achieved a precision of
better than 1\% for lithium~\cite{Amini2003a} and better than
0.1\% for sodium~\cite{Ekstrom1995a}, but this far-field
arrangement is rather demanding in itself and it is strongly
impeded for large molecules because it requires a collimation and
a brilliance beyond those of existing molecular beams.

On the other hand, it has been shown, that near-field
interferometry is well-suited for less intense sources of large
objects ~\cite{Clauser1997a} and Talbot-Lau deflectometry was
recently employed to measure the static polarizability of
fullerenes~\cite{Berninger2007a} as well as of porphyrins and
porphyrin derivatives~\cite{Deachapunya2007a}.

All earlier interferometric polarizability experiments relied on the
application of high static electric fields to shift the
atomic~\cite{Ekstrom1995a,Amini2003a} or
molecular~\cite{Berninger2007a} matter wave phase in proportion to
the particle's {\em static} polarizability.

Measurements of {\em optical (AC)} polarizabilities of clusters larger
than diatomic molecules have not been available until a few years ago
\cite{Ballard2000a}. As of today only very few methods have been
explored and exclusively demonstrated with fullerenes: Ballard et
al.~were the first to exploit the recoil imparted on a C$_{60}$ beam
when it crossed an intense standing light wave at 1064\,nm
\cite{Ballard2000a}. Their detector was a position sensitive
time-of-flight mass spectrometer which allowed to retrieve the
polarizability at 1064\,nm from a classical beam broadening.

In our present work we exploit near-field quantum diffraction in
a Kapitza-Dirac-Talbot-Lau configuration~\cite{Gerlich2007a} for
retrieving the optical polarizability of both C$_{70}$ and
C$_{60}$ from the evolution of the quantum fringe visibility as a
function of the diffracting laser power. This scheme offers a
higher throughput than in far-field diffraction experiments. Even
in the pure quantum regime it is in principal scalable to masses
beyond 10,000\,amu and it can thus yield interesting information
complementary to optical spectroscopy experiments.

\section{Setup}\label{sec:1}
Our new interferometer differs from an established earlier
version~\cite{Brezger2002a} in that it consists of two material
gratings and one central optical phase grating~\cite{Gerlich2007a}.
The first grating G$_1$ prepares the necessary coherence of the only
weakly collimated molecular beam. At the second grating G$_2$, which
is realized by a standing laser light wave at 532\,nm, diffraction
is based on the optical dipole potential
\begin{equation}
U(x,z)=-\frac{1}{2}\alpha_{L} E^2(x,z,t) , \label{dipolepotential}
\end{equation}
with the {\em scalar optical polarizability} $\alpha_{L}$ and the
electric field of the focused laser light wave $E$. Near-field
matter-wave interference, according to the Talbot effect, leads to
self-imaging~\cite{Clauser1994a,Dubetsky1997a,Hornberger2004a} of
the complex transmission function of G$_{2}$ into a periodic
molecular density distribution at the position of the third
grating G$_3$. The latter acts as a detector screen: shifting
this mask across the periodic interference pattern and detecting
the transmitted molecular flux then allows to reveal the
molecular interference pattern.

\begin{figure}[b]
 \includegraphics[width=1.00\columnwidth]{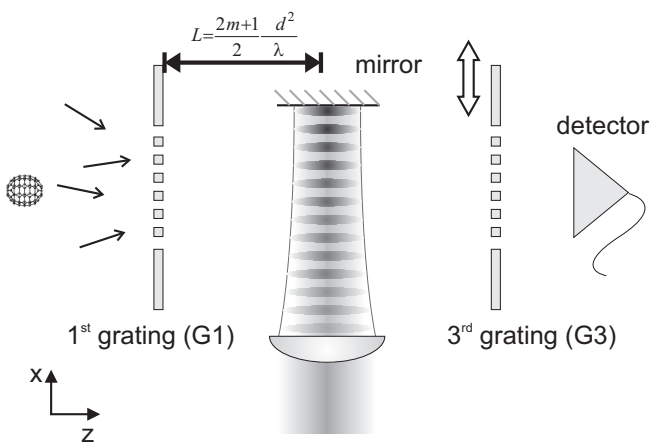}
 \caption{\label{fig:setup}
 Setup of the Kapitza-Dirac-Talbot-Lau interferometer: it consists of two
 material nanostructures and
 and one standing light wave. For detection, the third grating is shifted in
 20\,nm steps over the molecular pattern. The transmitted molecules
 are ionized and counted in an electron multiplier.}
\end{figure}

We designate our new setup as a Kapitza-Dirac-Talbot-Lau (KDTL)
interferometer~\cite{Gerlich2007a}, as it combines the virtue of
diffraction at standing light
waves~\cite{Martin1988a,Kapitza1933a,Freimund2001a,Nairz2001a}
with the Talbot-Lau concept. The central optical grating becomes
very important and useful for massive and highly polarizable
particles as well as for small grating openings: For material
gratings the influence of the van der Waals potential at the
second grating would increase with $\alpha$ and with shrinking
distance between the molecules and the grating walls. In
\cite{Grisenti1999a,Bruehl2002a} it has been shown that the
particle-wall interaction adds a velocity and position dependent
phase on the molecular wave function. In interferometry with
material gratings this imposes a rather demanding experimental
constraint, since high fringe contrasts can only be obtained in
very narrow velocity bands, typically of the order of $\Delta v/v
\le 1\%$ ~\cite{Brezger2002a,Gerlich2007a}. This requirement can
however be enormously alleviated by employing optical instead
of material gratings. It turns out to be sufficient to have an
optical diffraction grating in the center of the
KDTL-interferometer alone: The first and third material gratings
serve as absorptive masks for the preparation of coherence (G$_1$)
and the spatially resolved and parallel detection (G$_3$) of the
wide molecular beam. Additional phases which are imprinted onto
the molecules by these structures, play no significant role for
the formation of the interference pattern.

The SiN$_x$ material gratings have a period of d=266\,nm, a
thickness of 190\,nm and an open fraction of f=0.42. The latter is
the ratio between slit opening and grating period. The nanomasks
were manufactured by T. Savas (MIT, Cambridge, USA) with a period
homogeneity of better than $\Delta$d$\,\le$\,1\,\AA \,across the
entire width of the molecular beam. In the KDTLI setup G$_1$ and
G$_3$ are  separated by a distance of 210\,mm as sketched in
Fig.~1. The center of the standing light-wave is positioned with
an uncertainty of less than $20\,\mu$m in the middle between
these two masks. The light is derived from a green single-mode
laser (Coherent Verdi 10\,W at 532\,nm), tightly focused by a
cylindrical lens and back reflected from a planar mirror inside
the vacuum chamber.

The fullerene beam is generated by an effusive source and passes the
standing light wave in a distance of less than 300\,$\mu$m from the
mirror surface, where the laser beam is already focused to
$w_{x}=20\,\mu m$ and $w_{y}=900\,\mu m$. The molecules are collimated
by two slits to limit the angle of incidence onto the standing
light-wave to $90\pm\,0.02$\,degrees. This ensures that no molecule
will average nodes and an anti-nodes of the light grating.

The fullerenes are detected behind the interferometer by either
laser ionization, for C$_{70}$, or electron impact ionization
quadrupole mass spectrometry for C$_{60}$. The spatial resolution of
the detector is provided by the third grating which is mounted on a
piezo-electric stage. It can be shifted with a repeatability of
$\sim 10\,$nm. The material gratings are mounted on rotation stages
in order to allow alignment around the vertical axis and around the
molecular beam axis with better than 300\,$\mu$rad precision.
The velocity of the molecules is controlled by a gravitational velocity
selection scheme \cite{Brezger2002a}.

Interference in the Kapitza-Dirac-Talbot-Lau interferometer occurs
at de Broglie wavelengths which fulfill the KDTL-condition
\begin{equation}
L_{KDTL}= \frac{(2m+1)}{2} \frac{d^2}{\lambda_{dB}} ,
\label{optcrit}
\end{equation}
where m is an integer number. This condition, which is valid for
weak laser intensities ($\Phi_{max} < 3.05$, s. below) reveals
that the interference maximum recurs with integer multiples of the
Talbot period: $L_T=d^2/\lambda_{dB}$, but shifted by half a
period in comparison to the earlier Talbot-Lau
interferometer~\cite{Brezger2002a}. For fullerenes at velocities
between 100\,m/s and 190\,m/s our KDTLI-setup is operating
between the fourth and the eighth Talbot diffraction order.

\section{Theoretical model}\label{sec:2}
The conservative interaction of the molecule with the electric
field of the standing light-wave is described by the dipole
potential of Eq.~\ref{dipolepotential}. The molecules thus
acquire a position-dependent phase which enters the transmission
function $t_{2}(x)$ of G$_2$:
\begin{align}
t_{2}(x)=&\exp\left(-\frac{i}{\hbar}\int U(x,z(t),t)dt\right) \notag \\
=&\exp\left
(i\Phi_{max}\sin^2\left(\frac{2\pi}{\lambda_{L}}x\right)\right) ,
\end{align}
with a maximal phase shift at the anti-nodes of
\begin{equation}
\Phi_{max}=\frac{8\sqrt{2\pi}\alpha_{L}}{\hbar c w_{y}v_{z}}P ,
\end{equation}
where $\alpha_{L}$ is the real part of the optical polarizability
at the laser wavelength $\lambda_{L}$ and $P$ the light
power~\cite{Nairz2001a}. We further include the possibility of
photon absorption, which can lead to additional transverse
momentum kicks imparted onto the molecules. The absorption
process is governed by Poissonian statistics and the mean number
of absorbed photons in an anti-node of the standing wave is
\begin{align}
n_{0}=\frac{\sigma_{abs}}{\hbar \omega_{L}}\int I(x,z(t),t) dt =
\frac{8\sigma_{abs}\lambda_{L}}{\sqrt{2\pi}h c w_{y} v_{z}}P .
\end{align}
Here, $\sigma_{abs}$ is the absorption cross section at the laser
wavelength, determined  by the imaginary part of the optical
polarizability at the laser frequency $\omega_{L}$ and  $I$ is the
light intensity. The expected interference fringe contrast at
fixed molecule velocity is then given by~\cite{Gerlich2007a}
\begin{align}
V=&\, 2 \left( \frac{\sin(\pi f)}{\pi f}\right)^2 \exp(-\xi_{abs})
\frac{\xi_{coh}-\xi_{abs}}{\xi_{coh}+\xi_{abs}} \notag \\
&J_{2}\left(-\sgn(\xi_{coh} +
\xi_{abs})\sqrt{\xi_{coh}^2-\xi_{abs}^2} \right)
,\label{KDcontrast}
\end{align}
where
\begin{equation}
 \xi_{coh}=\Phi_{max} \sin \left(\pi \frac{L}{L_{T}} \right) ,
\end{equation}
accounts for the coherent interaction and
\begin{equation}
 \xi_{abs}=n_{0} \sin^2 \left( \frac{\pi L}{2 L_{T}}\right) ,
\end{equation}
for the incoherent scattering of photons. Here $L$ is the distance
between the gratings and $J_{2}$ is the Bessel function of the
second kind.

\section{Polarizability measurements}\label{sec:3}
As can be seen from the equations above, the molecular {\em
optical scalar polarizability $\alpha_{L}$} enters directly the
interference visibility of the KDTL-interferometer, through the
maximum phase shift $\Phi_{max}$. We can therefore exploit the
KDTLI to determine $\alpha_{L}$ when the resonant absorption
cross section $\sigma_{abs}$ is known from separate experiments,
such as gas-phase spectroscopy. In the following we compare the
experimentally observed dependence of the interference contrast
on the varying laser power and on the molecular velocity with the
theoretical model from section~\ref{sec:2}.

\begin{figure}
 \includegraphics[width=1.00\columnwidth]{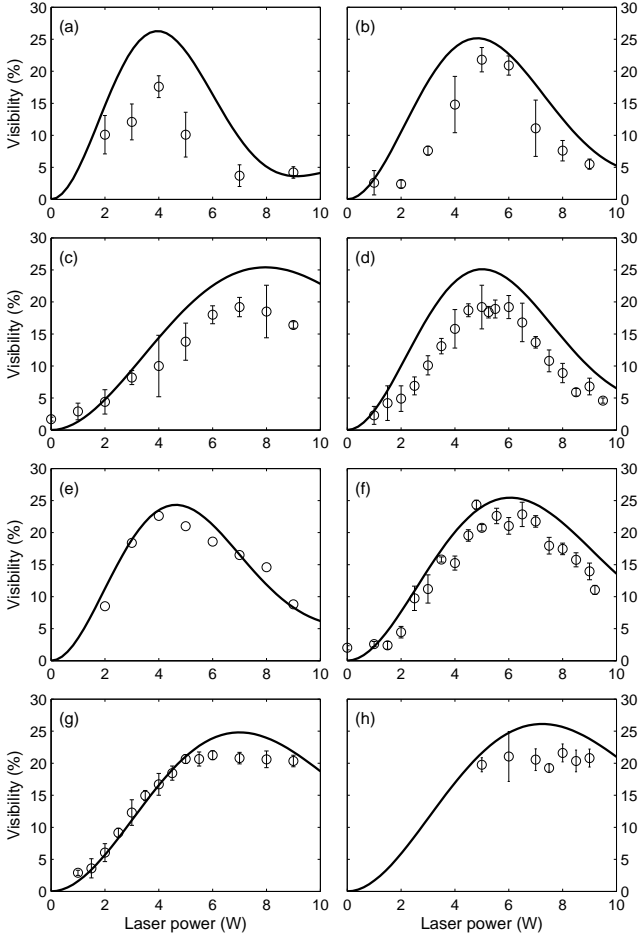}
 \caption{\label{fig:C70} Variation of the C$_{70}$ interference
 contrast with the diffracting laser power.
  The velocity distributions in these experiments were characterized
  by the following most probable velocities $v_m$ and velocity
  spreads $\Delta v$ (standard deviation): a) $v_m$=99.7\,m/s, $\Delta
  v$=18.3\,m/s, b) $v_m$=117.3\,m/s, $\Delta v$=14.4\,m/s, c) $v_m$=196.7\,m/s, $\Delta v$=39.5\,m/s,
  d) $v_m$= 124.6\,m/s, $\Delta v$= 22.8\,m/s, e) $v_m$= 114.4\,m/s, $\Delta v$= 18.8\,m/s,
  f) $v_m$= 152.7\,m/s, $\Delta v$= 24.8\,m/s, g) $v_m$= 171.2\,m/s, $\Delta v$= 28.8\,m/s,
  h) $v_m$= 179.9\,m/s, $\Delta v$= 33.5\,m/s.
 The experimental data are fitted with a single common parameter $\alpha_L (532\,\mathrm{nm})$ using equation
\eqref{KDcontrast}. The error bars on the data are standard
 deviations taken from three consecutive recordings with the same settings.  Figure e)
 has no error bar since only one data set was taken.} \label{C70interference}
\end{figure}

The interference patterns of C$_{70}$ are recorded for eight
different velocity distributions and for up to twenty different
laser powers $P$. For each sinusoidal interference curve we
extract the experimental fringe visibility $V=(S_{\rm max}- S_{\rm
min})/(S_{\rm max}+ S_{\rm min})$ and plot it versus the
diffracting laser power, as shown in Fig.~\ref{C70interference}.
We then fit the resulting curves with equation~\eqref{KDcontrast},
using the polarizability as a single free parameter, common for
all laser powers and velocities. Given the value for the optical
absorption cross section~\cite{Coheur1996a,Hornberger2005a}
$\sigma_{abs}(532\,nm)= 2.1\times 10^{-17}$cm$^{-2}$ and including
the experimentally determined velocity distributions in our
theoretical model we then determine the scalar optical
polarizability to be $\alpha_{L} (532\,\mathrm{nm})=117\pm
14$\,\AA$^3$.

This value is consistent with earlier results obtained from the
dielectric response of thin fullerene
films~\cite{Eklund1995a,Sohmen1992a}. It is also in good
agreement with measurements of the gas-phase static
polarizability which has been determined to be
$\alpha=102\,\pm\,14$\AA$^3$ using Stark
deflectometry~\cite{Compagnon2001c} and
$\alpha=108.5\,\pm\,2.0\pm\,6.2$\AA$^3$ by Talbot-Lau
deflectometry~\cite{Berninger2007a}. The optical polarizability
is expected to be higher than the static value since the incident
light approaches (still from far away) allowed optical transitions
at 532\,nm. A comparison of our work with literature values for
fullerene polarizabilities at 532\,nm is shown in
Table~\ref{tabalpha}.
\begin{table}
\caption{Optical polarizability of the fullerenes at 532\,nm: The
dielectric function of the thin film measurements is converted to
a polarizability using a lattice constant of $a=14.17$\,\AA\, for
C$_{60}$ and $a=15.2$\,\AA\, for C$_{70}$\cite{Sohmen1992a}.}

\label{tabalpha}
\begin{tabular}{lcccc}
\hline\noalign{\smallskip}
  & thin films \cite{Eklund1995a} & EELS \cite{Sohmen1992a} & theory ~\cite{Ruud2001a} & this work \\
\noalign{\smallskip}\hline\noalign{\smallskip}
C$_{60}$ & 90 \AA$^3$  & 98.2 \AA$^3$ & 80.6 &90(11)\AA$^3$  \\
C$_{70}$ & 118.4 \AA$^3$ &  122.6 \AA$^3$ & - &117(14)\AA$^3$ \\
\noalign{\smallskip}\hline
\end{tabular}
\end{table}

The error bars in our experiment have various different contributions:
Firstly, the absolute calibration of our present optical power meter
is good to within 10\%. This is a systematic uncertainty in the power
measurement, but the reproducibility and linearity in the reading are
better than 5\%. Secondly, the vertical waist $w_y$ of the laser beam
enters linearly in the available intensity. In our present
measurements it could been measured with an accuracy of 5\%. In
contrast to that, the horizontal laser waist (20$\mu$m) can only be
measured with an estimated accuracy of 20\%, but its uncertainty
cancels out in first order, since a larger waist is associated with a
lower intensity but also with a longer transition time for the
molecules. Thirdly, the knowledge of the optical absorption cross
section is relevant for the numerical fit of the data. However it
enters very weakly: even an uncertainty of 50\% in $\sigma_{abs}$
results only in a $3\%$ variation of the fit value for the
polarizability.  Fourthly, there is a statistical error, as can be
seen from the scatter of the data points. Taking a 'worst case
scenario' by separately fitting the upper and the lower envelope of
all error bars in the data distribution, we see a maximum uncertainty
of 2.3\,\AA$^3$.  Systematic deviations of the interferogram from an
ideal fringe contrast, which can be caused by day-to-day variations of
the laboratory noise background, enter very weakly into the fit value.
All experimental uncertainties mentioned above contribute
independently and sum up to a total uncertainty of 12\%.

Most parameters can clearly be improved in future experiments:
With an improved detection efficiency the width of the velocity
distribution can probably still be reduced by an order of
magnitude. The measurement of the laser power and waist may still
be improved by a factor of three using commercially available
sensors.

\begin{figure}[t]
 \includegraphics[width=1.00\columnwidth]{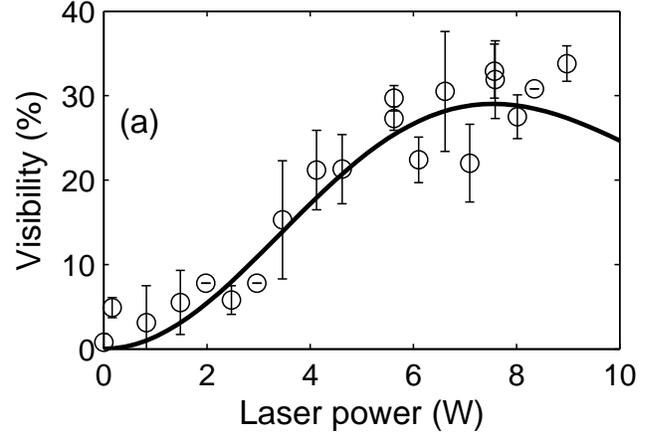}
 \caption{Variation of the C$_{60}$ interference contrast
 with the power of the diffracting laser light-wave. The
 experimental points are modeled with equation~\eqref{KDcontrast}
 and $\alpha_{L}$ as a free parameter after choosing $\sigma_{abs}$ from the literature. The error bars correspond
 to the standard deviation of three consecutive measurements.}
 \label{C60interference}
\end{figure}

In Fig.~\ref{C60interference} we show the power dependence of the
fringe contrast of $C_{60}$ for a mean molecular velocity of
153\,m/s and a standard deviation of $\Delta v/v=0.3$. The larger
error bars here are caused by the lower detection efficiency of
$C_{60}$ in comparison to $C_{70}$. We again use
equation~\ref{KDcontrast} to fit the experimental data points and
evaluate, for a given absorption cross section of
$\sigma_{abs}=3.2\times10^{-18}$cm$^2$
\cite{Coheur1996a,Ferrante2003a}, the dynamic polarizability of
C$_{60}$ at 532\,nm to be $\alpha_{L}= 91 (11)$\,\AA$^3$. This
value is in good agreement with previous measurements, as shown in
Table~\ref{tabalpha} and also consistent with the static
polarizability values which were determined by Stark-deflectometry
($\alpha=76.5\,\pm\,8.0$\AA$^3$~\cite{Antoine1999a}) and by
interferometric Talbot-Lau deflectometry
($\alpha=88.9\,\pm\,0.09\pm\,5.1$\AA$^3$~\cite{Berninger2007a}).

In conclusion we have used a Kapitza-Dirac-Talbot-Lau
interferometer to determine the scalar optical polarizability of
both Fullerenes C$_{60}$ and C$_{70}$. Our method can be extended
to all volatile molecules that may be subjected to
KDTL-interferometry. In the present setup this can cover a wide
class of effusive beams in principle with molecular masses up to
10,000\,amu.

Combined with an independent evaluation of optical absorption
cross sections the new technique can be an interesting complement
to existing methods~\cite{Ballard2000a,Nairz2001a,Bonin1997a} and
it will provide new data for a large set of molecules as an
experimental benchmark for new computational molecular models.

\section*{Acknowledgements}
This work has been supported by the Austrian Science Foundation
(FWF), within the SFB projects F1505 and F1512 and START Y177.
K.H. acknowledges support within the Emmy Noether program by the
German Science Foundation DFG.


\begin{thebibliography}{36}
\expandafter\ifx\csname
natexlab\endcsname\relax\def\natexlab#1{#1}\fi
\expandafter\ifx\csname bibnamefont\endcsname\relax
  \def\bibnamefont#1{#1}\fi
\expandafter\ifx\csname bibfnamefont\endcsname\relax
  \def\bibfnamefont#1{#1}\fi
\expandafter\ifx\csname citenamefont\endcsname\relax
  \def\citenamefont#1{#1}\fi
\expandafter\ifx\csname url\endcsname\relax
  \def\url#1{\texttt{#1}}\fi
\expandafter\ifx\csname
urlprefix\endcsname\relax\def\urlprefix{URL }\fi
\providecommand{\bibinfo}[2]{#2}
\providecommand{\eprint}[2][]{\url{#2}}

\bibitem[1]{Shull1969a}
\bibinfo{author}{\bibfnamefont{C.~G.} \bibnamefont{Shull}},
  \bibinfo{journal}{Phys. Rev.} \textbf{\bibinfo{volume}{179}},
  \bibinfo{pages}{752} (\bibinfo{year}{1969}).

\bibitem[2]{Mollenstedt1959a}
\bibinfo{author}{\bibfnamefont{G.}~\bibnamefont{Moellenstedt}}
  \bibnamefont{and} \bibinfo{author}{\bibfnamefont{C.~Z.}
  \bibnamefont{Joensson}}, \bibinfo{journal}{Z. Phys.}
  \textbf{\bibinfo{volume}{155}}, \bibinfo{pages}{472} (\bibinfo{year}{1959}).

\bibitem[3]{Carnal1991a}
\bibinfo{author}{\bibfnamefont{O.}~\bibnamefont{Carnal}} \bibnamefont{and}
  \bibinfo{author}{\bibfnamefont{J.}~\bibnamefont{Mlynek}},
  \bibinfo{journal}{Phys. Rev. Lett.} \textbf{\bibinfo{volume}{66}},
  \bibinfo{pages}{2689} (\bibinfo{year}{1991}).

\bibitem[4]{Keith1991a}
\bibinfo{author}{\bibfnamefont{D.~W.} \bibnamefont{Keith}},
  \bibinfo{author}{\bibfnamefont{C.~R.} \bibnamefont{Ekstrom}},
  \bibinfo{author}{\bibfnamefont{Q.~A.} \bibnamefont{Turchette}},
  \bibnamefont{and} \bibinfo{author}{\bibfnamefont{D.~E.}
  \bibnamefont{Pritchard}}, \bibinfo{journal}{Phys. Rev. Lett.}
  \textbf{\bibinfo{volume}{66}}, \bibinfo{pages}{2693} (\bibinfo{year}{1991}).

\bibitem[5]{Rauch2000a}
\bibinfo{author}{\bibfnamefont{H.}~\bibnamefont{Rauch}} \bibnamefont{and}
  \bibinfo{author}{\bibfnamefont{A.}~\bibnamefont{Werner}},
  \emph{\bibinfo{title}{Neutron Interferometry: Lessons in Experimental Quantum
  Mechanics}} (\bibinfo{publisher}{Oxford Univ. Press}, Oxford \bibinfo{year}{2000}).

\bibitem[6]{Price1996a}
\bibinfo{author}{\bibfnamefont{D.~L.} \bibnamefont{Price}},
  \emph{\bibinfo{title}{Neutron Scattering, Part B, 23 Experimental Methods in
  the Physical Sciences}} (\bibinfo{publisher}{Academic Press Inc.},
  New York
  \bibinfo{year}{1996}).

\bibitem[7]{Tonomura1987a}
\bibinfo{author}{\bibfnamefont{A.}~\bibnamefont{Tonomura}},
  \bibinfo{journal}{Reviews of Modern Physics} \textbf{\bibinfo{volume}{59}},
  \bibinfo{pages}{639} (\bibinfo{year}{1987}).

\bibitem[8]{Cronin2007a}
\bibinfo{author}{\bibfnamefont{A.~D.} \bibnamefont{Cronin}},
  \bibinfo{author}{\bibfnamefont{J.}~\bibnamefont{Schmiedmayer}},
  \bibnamefont{and} \bibinfo{author}{\bibfnamefont{D.~E.}
  \bibnamefont{Pritchard}}, \bibinfo{journal}{Rev. Mod. Phys.}
  (\bibinfo{year}{2007}).

\bibitem[9]{Arndt1999a}
\bibinfo{author}{\bibfnamefont{M.}~\bibnamefont{Arndt}},
  \bibinfo{author}{\bibfnamefont{O.}~\bibnamefont{Nairz}},
  \bibinfo{author}{\bibfnamefont{J.}~\bibnamefont{Voss-Andreae}},
  \bibinfo{author}{\bibfnamefont{C.}~\bibnamefont{Keller}},
  \bibinfo{author}{\bibfnamefont{G.~V.} \bibnamefont{der Zouw}},
  \bibnamefont{and}
  \bibinfo{author}{\bibfnamefont{A.}~\bibnamefont{Zeilinger}},
  \bibinfo{journal}{Nature} \textbf{\bibinfo{volume}{401}},
  \bibinfo{pages}{680} (\bibinfo{year}{1999}).

\bibitem[10]{Berninger2007a}
\bibinfo{author}{\bibfnamefont{M.}~\bibnamefont{Berninger}},
  \bibinfo{author}{\bibfnamefont{A.}~\bibnamefont{St{\'e}fanov}},
  \bibinfo{author}{\bibfnamefont{S.}~\bibnamefont{Deachapunya}},
  \bibnamefont{and} \bibinfo{author}{\bibfnamefont{M.}~\bibnamefont{Arndt}},
  \bibinfo{journal}{Phys. Rev. A}  (\bibinfo{year}{2007}).

\bibitem[11]{Deachapunya2007a}
\bibinfo{author}{\bibfnamefont{S.}~\bibnamefont{Deachapunya}},
  \bibinfo{author}{\bibfnamefont{P.~J.} \bibnamefont{Fagan}},
  \bibinfo{author}{\bibfnamefont{A.~G.} \bibnamefont{Major}},
  \bibinfo{author}{\bibfnamefont{E.}~\bibnamefont{Reiger}},
  \bibinfo{author}{\bibfnamefont{H.}~\bibnamefont{Ritsch}},
  \bibinfo{author}{\bibfnamefont{A.}~\bibnamefont{Stefanov}},
  \bibinfo{author}{\bibfnamefont{H.}~\bibnamefont{Ulbricht}}, \bibnamefont{and}
  \bibinfo{author}{\bibfnamefont{M.}~\bibnamefont{Arndt}},
  \bibinfo{journal}{http://arxiv.org/abs/0708.1449}  (\bibinfo{year}{2007}).

\bibitem[12]{Gaussian2003a}
\bibinfo{author}{\bibfnamefont{M.}~\bibnamefont{Frisch}},
  \bibinfo{author}{\bibfnamefont{G.}~\bibnamefont{Trucks}},
  \bibinfo{author}{\bibfnamefont{H.}~\bibnamefont{Schlegel}},
  \bibinfo{author}{\bibfnamefont{G.}~\bibnamefont{Scuseria}},
  \bibinfo{author}{\bibfnamefont{M.}~\bibnamefont{Robb}},
  \bibinfo{author}{\bibfnamefont{J.}~\bibnamefont{Cheeseman}},
  \bibinfo{author}{\bibfnamefont{J.}~\bibnamefont{Montgomery}},
  \bibinfo{author}{\bibnamefont{Jr.}},
  \bibinfo{author}{\bibfnamefont{T.}~\bibnamefont{Vreven}},
  \bibinfo{author}{\bibfnamefont{K.}~\bibnamefont{Kudin}},
  \bibnamefont{et~al.}, \emph{\bibinfo{title}{Gaussian 03W, Version 6.0}}
  (\bibinfo{publisher}{Gaussian Inc.}, \bibinfo{address}{Pittsburgh, PA},
  \bibinfo{year}{2003}).

\bibitem[13]{Bonin1997a}
\bibinfo{author}{\bibfnamefont{K.}~\bibnamefont{Bonin}} \bibnamefont{and}
  \bibinfo{author}{\bibfnamefont{V.}~\bibnamefont{Kresin}},
  \emph{\bibinfo{title}{Electric-Dipole Polarizabilities of Atoms, Molecules
  and Clusters}} (\bibinfo{publisher}{World Scientific}, Singapore \bibinfo{year}{1997}),
  ISBN \bibinfo{isbn}{981-02-2493-1}.

\bibitem[14]{Antoine1999a}
\bibinfo{author}{\bibfnamefont{R.}~\bibnamefont{Antoine}},
  \bibinfo{author}{\bibfnamefont{P.}~\bibnamefont{Dugourd}},
  \bibinfo{author}{\bibfnamefont{D.}~\bibnamefont{Rayane}},
  \bibinfo{author}{\bibfnamefont{E.}~\bibnamefont{Benichou}},
  \bibinfo{author}{\bibfnamefont{M.}~\bibnamefont{Broyer}},
  \bibinfo{author}{\bibfnamefont{F.}~\bibnamefont{Chandezon}},
  \bibnamefont{and} \bibinfo{author}{\bibfnamefont{C.}~\bibnamefont{Guet}},
  \bibinfo{journal}{J. Chem. Phys.} \textbf{\bibinfo{volume}{110}},
  \bibinfo{pages}{9771 } (\bibinfo{year}{1999}).

\bibitem[15]{Amini2003a}
\bibinfo{author}{\bibfnamefont{J.~M.} \bibnamefont{Amini}} \bibnamefont{and}
  \bibinfo{author}{\bibfnamefont{H.}~\bibnamefont{Gould}},
  \bibinfo{journal}{Phys. Rev. Lett.} \textbf{\bibinfo{volume}{91}},
  \bibinfo{pages}{153001} (\bibinfo{year}{2003}).

\bibitem[16]{Ekstrom1995a}
\bibinfo{author}{\bibfnamefont{C.}~\bibnamefont{Ekstrom}},
  \bibinfo{author}{\bibfnamefont{J.}~\bibnamefont{Schmiedmayer}},
  \bibinfo{author}{\bibfnamefont{M.}~\bibnamefont{Chapman}},
  \bibinfo{author}{\bibfnamefont{T.}~\bibnamefont{Hammond}}, \bibnamefont{and}
  \bibinfo{author}{\bibfnamefont{D.}~\bibnamefont{Pritchard}},
  \bibinfo{journal}{Phys. Rev. A} \textbf{\bibinfo{volume}{51}},
  \bibinfo{pages}{3883} (\bibinfo{year}{1995}).

\bibitem[17]{Clauser1997a}
\bibinfo{author}{\bibfnamefont{J.}~\bibnamefont{Clauser}}, in
  \emph{\bibinfo{booktitle}{Experimental Metaphysics}}, edited by
  \bibinfo{editor}{\bibfnamefont{R.}~\bibnamefont{Cohen}},
  \bibinfo{editor}{\bibfnamefont{M.}~\bibnamefont{Horne}}, \bibnamefont{and}
  \bibinfo{editor}{\bibfnamefont{J.}~\bibnamefont{Stachel}}
  (\bibinfo{publisher}{Kluwer Academic}, Dordrecht \bibinfo{year}{1997}).

\bibitem[18]{Ballard2000a}
\bibinfo{author}{\bibfnamefont{A.}~\bibnamefont{Ballard}},
  \bibinfo{author}{\bibfnamefont{K.}~\bibnamefont{Bonin}}, \bibnamefont{and}
  \bibinfo{author}{\bibfnamefont{J.}~\bibnamefont{Louderback}},
  \bibinfo{journal}{J. Chem. Phys.} \textbf{\bibinfo{volume}{114}},
  \bibinfo{pages}{5732} (\bibinfo{year}{2000}).

\bibitem[19]{Nairz2001a}
\bibinfo{author}{\bibfnamefont{O.}~\bibnamefont{Nairz}},
  \bibinfo{author}{\bibfnamefont{B.}~\bibnamefont{Brezger}},
  \bibinfo{author}{\bibfnamefont{M.}~\bibnamefont{Arndt}}, \bibnamefont{and}
  \bibinfo{author}{\bibfnamefont{A.}~\bibnamefont{Zeilinger}},
  \bibinfo{journal}{Phys. Rev. Lett.} \textbf{\bibinfo{volume}{87}},
  \bibinfo{pages}{160401} (\bibinfo{year}{2001}).

\bibitem[20]{Gerlich2007a}
\bibinfo{author}{\bibfnamefont{S.}~\bibnamefont{Gerlich}},
  \bibinfo{author}{\bibfnamefont{L.}~\bibnamefont{Hackerm\"{u}ller}},
  \bibinfo{author}{\bibfnamefont{K.}~\bibnamefont{Hornberger}},
  \bibinfo{author}{\bibfnamefont{A.}~\bibnamefont{Stibor}},
  \bibinfo{author}{\bibfnamefont{H.}~\bibnamefont{Ulbricht}},
  \bibinfo{author}{\bibfnamefont{M.}~\bibnamefont{Gring}},
  \bibinfo{author}{\bibfnamefont{F.}~\bibnamefont{Goldfarb}},
  \bibinfo{author}{\bibfnamefont{T.}~\bibnamefont{Savas}},
  \bibinfo{author}{\bibfnamefont{M.}~\bibnamefont{M\"{u}ri}},
  \bibinfo{author}{\bibfnamefont{M.}~\bibnamefont{Mayor}}, and
  \bibinfo{author}{\bibfnamefont{M.}~\bibnamefont{Arndt}},
  \bibinfo{journal}{Nature Physics}, doi:10.1038/nphys701
  (\bibinfo{year}{2007}).

\bibitem[21]{Brezger2002a}
\bibinfo{author}{\bibfnamefont{B.}~\bibnamefont{Brezger}},
  \bibinfo{author}{\bibfnamefont{L.}~\bibnamefont{Hackerm{\"u}ller}},
  \bibinfo{author}{\bibfnamefont{S.}~\bibnamefont{Uttenthaler}},
  \bibinfo{author}{\bibfnamefont{J.}~\bibnamefont{Petschinka}},
  \bibinfo{author}{\bibfnamefont{M.}~\bibnamefont{Arndt}}, \bibnamefont{and}
  \bibinfo{author}{\bibfnamefont{A.}~\bibnamefont{Zeilinger}},
  \bibinfo{journal}{Phys. Rev. Lett.} \textbf{\bibinfo{volume}{88}},
  \bibinfo{pages}{100404} (\bibinfo{year}{2002}).

\bibitem[22]{Clauser1994a}
\bibinfo{author}{\bibfnamefont{J.~F.} \bibnamefont{Clauser}} \bibnamefont{and}
  \bibinfo{author}{\bibfnamefont{S.}~\bibnamefont{Li}}, \bibinfo{journal}{Phys.
  Rev. A} \textbf{\bibinfo{volume}{49}}, \bibinfo{pages}{R2213}
  (\bibinfo{year}{1994}).

\bibitem[23]{Dubetsky1997a}
\bibinfo{author}{\bibfnamefont{B.}~\bibnamefont{Dubetsky}} \bibnamefont{and}
  \bibinfo{author}{\bibfnamefont{P.~R.} \bibnamefont{Berman}}, in
  \emph{\bibinfo{booktitle}{Atom Interferometry}}, edited by
  \bibinfo{editor}{\bibfnamefont{P.~R.} \bibnamefont{Berman}}
  (\bibinfo{publisher}{Academic Press}, \bibinfo{address}{San Diego}
  \bibinfo{year}{1997}).

\bibitem[24]{Hornberger2004a}
\bibinfo{author}{\bibfnamefont{K.}~\bibnamefont{Hornberger}},
  \bibinfo{author}{\bibfnamefont{J.~E.} \bibnamefont{Sipe}}, \bibnamefont{and}
  \bibinfo{author}{\bibfnamefont{M.}~\bibnamefont{Arndt}},
  \bibinfo{journal}{Phys. Rev. A} \textbf{\bibinfo{volume}{70}},
  \bibinfo{pages}{53608} (\bibinfo{year}{2004}).

\bibitem[25]{Martin1988a}
\bibinfo{author}{\bibfnamefont{P.~J.} \bibnamefont{Martin}},
  \bibinfo{author}{\bibfnamefont{B.~G.} \bibnamefont{Oldaker}},
  \bibinfo{author}{\bibfnamefont{A.~H.} \bibnamefont{Miklich}},
  \bibnamefont{and} \bibinfo{author}{\bibfnamefont{D.~E.}
  \bibnamefont{Pritchard}}, \bibinfo{journal}{Phys. Rev. Lett.}
  \textbf{\bibinfo{volume}{60}}, \bibinfo{pages}{515} (\bibinfo{year}{1988}).

\bibitem[26]{Kapitza1933a}
\bibinfo{author}{\bibfnamefont{P.~L.} \bibnamefont{Kapitza}} \bibnamefont{and}
  \bibinfo{author}{\bibfnamefont{P.~A.~M.} \bibnamefont{Dirac}},
  \bibinfo{journal}{Proc. Camb. Philos. Soc.} \textbf{\bibinfo{volume}{29}},
  \bibinfo{pages}{297 } (\bibinfo{year}{1933}).

\bibitem[27]{Freimund2001a}
\bibinfo{author}{\bibfnamefont{D.~L.} \bibnamefont{Freimund}}, 
 K. Aflatooni, and H. Batelaan, Nature \textbf{413}, 142 (2001).

Ph.D. thesis,
  \bibinfo{school}{University of Nebraska} (\bibinfo{year}{2003}).

\bibitem[28]{Grisenti1999a}
\bibinfo{author}{\bibfnamefont{R.~E.} \bibnamefont{Grisenti}},
  \bibinfo{author}{\bibfnamefont{W.}~\bibnamefont{Sch{\"o}llkopf}},
  \bibinfo{author}{\bibfnamefont{J.~P.} \bibnamefont{Toennies}},
  \bibinfo{author}{\bibfnamefont{G.~C.} \bibnamefont{Hegerfeldt}},
  \bibnamefont{and}
  \bibinfo{author}{\bibfnamefont{T.}~\bibnamefont{K{\"o}hler}},
  \bibinfo{journal}{Phys. Rev. Lett.} \textbf{\bibinfo{volume}{83}},
  \bibinfo{pages}{1755} (\bibinfo{year}{1999}).

\bibitem[29]{Bruehl2002a}
\bibinfo{author}{\bibfnamefont{R.}~\bibnamefont{Br{\"u}hl}},
  \bibinfo{author}{\bibfnamefont{P.}~\bibnamefont{Fouquet}},
  \bibinfo{author}{\bibfnamefont{R.~E.} \bibnamefont{Grisenti}},
  \bibinfo{author}{\bibfnamefont{J.~P.} \bibnamefont{Toennies}},
  \bibinfo{author}{\bibfnamefont{G.~C.} \bibnamefont{Hegerfeldt}},
  \bibinfo{author}{\bibfnamefont{T.}~\bibnamefont{K{\"o}hler}},
  \bibinfo{author}{\bibfnamefont{M.}~\bibnamefont{Stoll}}, \bibnamefont{and}
  \bibinfo{author}{\bibfnamefont{C.}~\bibnamefont{Walter}},
  \bibinfo{journal}{Europhys. Lett.} \textbf{\bibinfo{volume}{59}},
  \bibinfo{pages}{357 } (\bibinfo{year}{2002}).

\bibitem[30]{Coheur1996a}
\bibinfo{author}{\bibfnamefont{P.~F.} \bibnamefont{Coheur}},
  \bibinfo{author}{\bibfnamefont{M.}~\bibnamefont{Carleer}}, \bibnamefont{and}
  \bibinfo{author}{\bibfnamefont{R.}~\bibnamefont{Colin}}, \bibinfo{journal}{J.
  Phys. B: At. Mol. Opt. Phys.} \textbf{\bibinfo{volume}{29}},
  \bibinfo{pages}{4987 } (\bibinfo{year}{1996}).

\bibitem[31]{Hornberger2005a}
\bibinfo{author}{\bibfnamefont{K.}~\bibnamefont{Hornberger}},
  \bibinfo{author}{\bibfnamefont{L.}~\bibnamefont{Hackerm\"{u}ller}},
  \bibnamefont{and} \bibinfo{author}{\bibfnamefont{M.}~\bibnamefont{Arndt}},
  \bibinfo{journal}{Physical Review A} \textbf{\bibinfo{volume}{71}},
  \bibinfo{pages}{023601} (\bibinfo{year}{2005}).

\bibitem[32]{Eklund1995a}
\bibinfo{author}{\bibfnamefont{P.}~\bibnamefont{Eklund}},
  \bibinfo{author}{\bibfnamefont{A.}~\bibnamefont{Rao}},
  \bibinfo{author}{\bibfnamefont{Y.}~\bibnamefont{Wang}},
  \bibinfo{author}{\bibfnamefont{P.}~\bibnamefont{Zhou}},
  \bibinfo{author}{\bibfnamefont{K.}~\bibnamefont{Wang}},
  \bibinfo{author}{\bibfnamefont{J.}~\bibnamefont{Holden}},
  \bibinfo{author}{\bibfnamefont{M.}~\bibnamefont{Dresselhaus}},
  \bibnamefont{and}
  \bibinfo{author}{\bibfnamefont{G.}~\bibnamefont{Dresselhaus}},
  \bibinfo{journal}{Thin Solid films} \textbf{\bibinfo{volume}{257}},
  \bibinfo{pages}{211} (\bibinfo{year}{1995}).

\bibitem[33]{Sohmen1992a}
\bibinfo{author}{\bibfnamefont{E.}~\bibnamefont{Sohmen}},
  \bibinfo{author}{\bibfnamefont{J.}~\bibnamefont{Fink}}, \bibnamefont{and}
  \bibinfo{author}{\bibfnamefont{W.}~\bibnamefont{Kr{\"a}tschmer}},
  \bibinfo{journal}{Z. Phys. B} \textbf{\bibinfo{volume}{86}},
  \bibinfo{pages}{87} (\bibinfo{year}{1992}).

\bibitem[34]{Compagnon2001c}
\bibinfo{author}{\bibfnamefont{I.}~\bibnamefont{Compagnon}},
  \bibinfo{author}{\bibfnamefont{R.}~\bibnamefont{Antoine}},
  \bibinfo{author}{\bibfnamefont{M.}~\bibnamefont{Broyer}},
  \bibinfo{author}{\bibfnamefont{P.}~\bibnamefont{Dugourd}},
  \bibinfo{author}{\bibfnamefont{J.}~\bibnamefont{Lerme}}, \bibnamefont{and}
  \bibinfo{author}{\bibfnamefont{D.}~\bibnamefont{Rayane}},
  \bibinfo{journal}{Phys. Rev. A} \textbf{\bibinfo{volume}{64}},
  \bibinfo{pages}{025201} (\bibinfo{year}{2001}).

\bibitem[35]{Ruud2001a}
\bibinfo{author}{\bibfnamefont{K.}~\bibnamefont{Ruud}},
  \bibinfo{author}{\bibfnamefont{D.}~\bibnamefont{Jonsson}}, \bibnamefont{and}
  \bibinfo{author}{\bibfnamefont{P.~R.} \bibnamefont{Taylor}},
  \bibinfo{journal}{J. Chem. Phys.} \textbf{\bibinfo{volume}{114}},
  \bibinfo{pages}{4331} (\bibinfo{year}{2001}).

\bibitem[36]{Ferrante2003a}
\bibinfo{author}{\bibfnamefont{C.}~\bibnamefont{Ferrante}},
  \bibinfo{author}{\bibfnamefont{R.}~\bibnamefont{Signorini}},
  \bibinfo{author}{\bibfnamefont{A.}~\bibnamefont{Feis}}, \bibnamefont{and}
  \bibinfo{author}{\bibfnamefont{R.}~\bibnamefont{Bozio}},
  \bibinfo{journal}{Photochem. Photobiol. Sci.} \textbf{\bibinfo{volume}{2}},
  \bibinfo{pages}{801} (\bibinfo{year}{2003}).

\end{thebibliography}

\end{document}